\documentclass{optica-article}

\journal{opticajournal} 

\articletype{Research Article}

\usepackage{lineno}
\usepackage{tabu}
\usepackage{multirow}
\usepackage{verbatim}
\usepackage{subcaption}
\usepackage{booktabs}

\begin{document}

\title{Chip-Scale Transmitter Module for Real-Time Continuous-Variable QKD}

\author{Igor Servello\authormark{1,2}, Martin Hauer\authormark{1}, Moritz Baier\authormark{1}, Emmeran Sollner\authormark{1},  Peter Gleißner\authormark{1}, Sebastian Randel\authormark{2}, Ulrich Eismann\authormark{1}, Emanuel Eichhammer\authormark{1}, Imran Khan\authormark{1,*}}

\address{\authormark{1}KEEQuant GmbH, Gebhardtstraße 28, 90762 Fürth, Germany\\
\authormark{2}Institute of Photonics and Quantum Electronics (IPQ), Karlsruhe Institute of Technology (KIT), Karlsruhe, Germany}

\email{\authormark{*}imran.khan@keequant.com}


\begin{abstract*} 
Continuous-variable quantum key distribution (CV-QKD) enables secure communication over standard telecom infrastructure, but scaling is stalled by bulky, discrete optical hardware. We address this bottleneck by demonstrating a real-time CV-QKD system driven by a chip-scale hybrid transmitter using commercial telecom components. Combining a micro-optic external-cavity laser with a monolithic photonic integrated IQ modulator we enable secure secret-key generation over 102 km of optical fiber while reducing optical volume by 95\% relative to the commercial discrete-component counterpart. Moreover, real-time operation overcomes offline post-processing bottlenecks of experimental setups. This work bridges laboratory demonstrations and field-deployable technology for cost-effective quantum networks.
\end{abstract*}

\section{Introduction}
The recent advances in quantum computing have highlighted significant vulnerabilities in conventional cryptographic methods. Shor's algorithm~\cite{Shor1994} enables sufficiently powerful quantum computers to break many asymmetric cryptography schemes, notably RSA (Rivest–Shamir–Adleman)~\cite{RSA1978} and DH (Diffie–Hellman)~\cite{1055638}.  
More generally, the security of classical asymmetric cryptography relies on computational hardness assumptions based on current algorithmic theory; consequently, it remains vulnerable to future cryptography breakthroughs that could invalidate these assumptions. Therefore, current and future communications are at risk, as quantum computers or advances in algorithms could enable retroactive attacks on stored encrypted data, that is, "store-now, decrypt-later” attacks.

\begin{figure}[!ht]
\centering
\includegraphics[width=0.65\linewidth]{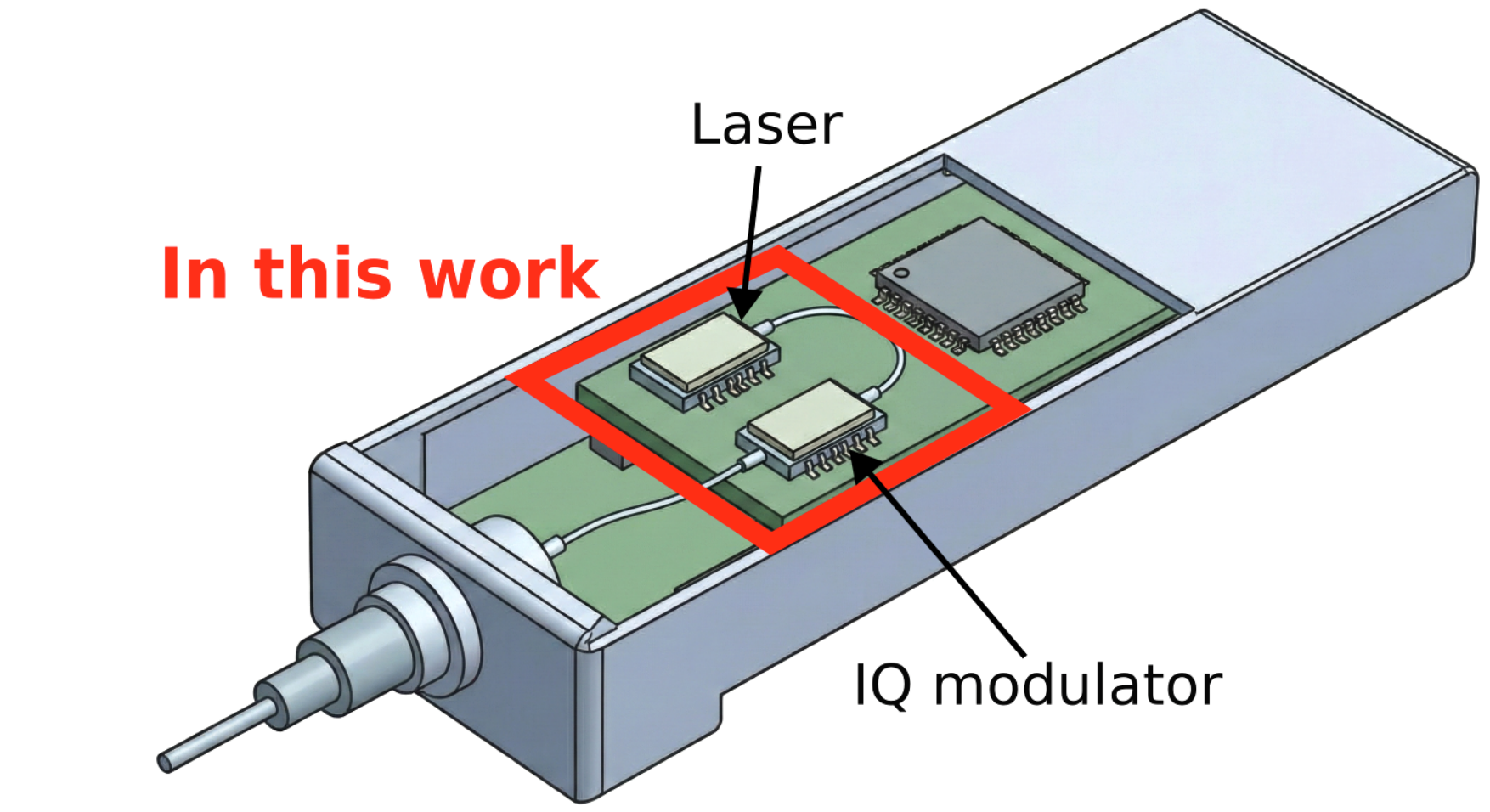}
\caption{Concept of a real-time chip-scale CV-QKD transmitter module. This work demonstrates the miniaturization of the optical components of the system (highlighted by the red square) and the automation of the CV-QKD pipeline. \textit{Parts of this figure were generated using the AI-assisted image generation tool Nano Banana.}}
\label{fig:integrated_alice}
\end{figure}

Quantum Key Distribution (QKD) is immune against these threats. QKD enables two distant parties to securely establish symmetric cryptographic keys under the assumption of an attacker with infinite (quantum) computational power, enabling true long-term security guaranteed by the principles of quantum mechanics. The first QKD protocol, BB84~\cite{BB84}, exploits the non-orthogonality of two polarization bases and the fundamental behavior of single photons. Following the BB84 protocol, a variety of QKD schemes have been proposed, exploiting different properties of quantum light and detection techniques \cite{PhysRevLett.67.661,PhysRevLett.68.3121,PhysRevLett.108.130503}. Among these, Continuous Variable QKD (CV-QKD) has emerged as a particularly promising approach \cite{Ralph_1999}.\\
In CV-QKD, information is encoded on the quadratures of coherent states of light, and the receiver uses coherent detection to measure these quadratures. The security of CV-QKD originates from the non-orthogonality of the quadratures of coherent states. Due to the Heisenberg uncertainty principle, both quadratures cannot be simultaneously measured with arbitrary precision. Any attempt to do so inevitably introduces noise, which enables the detection of eavesdropping and thus allows the implementation of QKD protocols.
Because CV-QKD encodes information onto the quadratures of coherent states, it is inherently compatible with standard coherent optical communication systems and can directly leverage their well-established tools and techniques. This compatibility allows the reuse of existing telecom infrastructure, hardware, and digital signal processing (DSP) algorithms, enabling the implementation of high-rate, cost-effective and scalable quantum key distribution systems that can operate alongside conventional optical networks. In this regards, recent works focusing on high symbol rates~\cite{Ren2021DemonstrationOH,Wang2018HighKR,Wang2020HighspeedGC,Wang2021SubGbpsKR,Wang2024HighKR}, long transmission distances~\cite{Hajomer2023LongdistanceCQ,Zhang2020LongDistanceCQ,Huang2015MaterialF,Qi2025LongDL}, and co-propagation of quantum signals with classical telecom channels~\cite{Hajomer2025CoexistenceOC,Kawakami2025CoexistenceDA,Geng2021CoexistenceOQ,Wang2023FieldTO,Shao2025IntegrationOC,Kleis2019ExperimentalIO,Gavignet2023CopropagationO6,Melgar2025CoexistenceOC,Eriksson2019CrosstalkIO,Kawakami2024NoguardbandIO,Honz2022FirstDO}, showcase the potential for widespread adoption and practical deployment.

The analogy to classical coherent telecommunications suggests that CV-QKD can follow the same path: miniaturization and cost reduction by means of photonic integrated circuits (PICs).
Significant progress has been made towards integrating key components on-chip, including lasers~\cite{Li2023}, IQ modulators~\cite{Aldama2025}, variable optical attenuators (VOAs)~\cite{Bian2024a}, balanced detectors~\cite{Pietri2024,Bian2024b,Hajomer2024}, and combined IQ-modulator/coherent-receiver architectures~\cite{Zhang2019,Hajomer2025b,Ng2025,Liu2025, 11263038}. Despite these advances, the demonstrations to our knowledge have achieved only partially miniaturized modules, as they still rely on essential components based on bulk optics, which prevented the realization of a scalable module. Furthermore, most implementations remain confined to laboratory environments, and rely on offline post-processing, limiting the deployability of the system to real networks.

In this work, we report the first demonstration of a CV-QKD transmitter composed entirely of chip-scale components, combining miniaturized laser source and modulator, as illustred in Fig.~\ref{fig:integrated_alice}. The result is a scalable CV-QKD transmitter module built entirely from telecom-grade, compact components. The QKD system employing the chip-scale transmitter operates in real time, and features fully automated field-deployable operation. The system delivers high performance, with validated key generation over metropolitan-scale optical fiber links. This work represents a key step from laboratory-scale experiments toward practical, scalable quantum-secure networks.

\section{Real-time CV-QKD System}
\begin{figure}[]
\centering
\includegraphics[width=\linewidth]{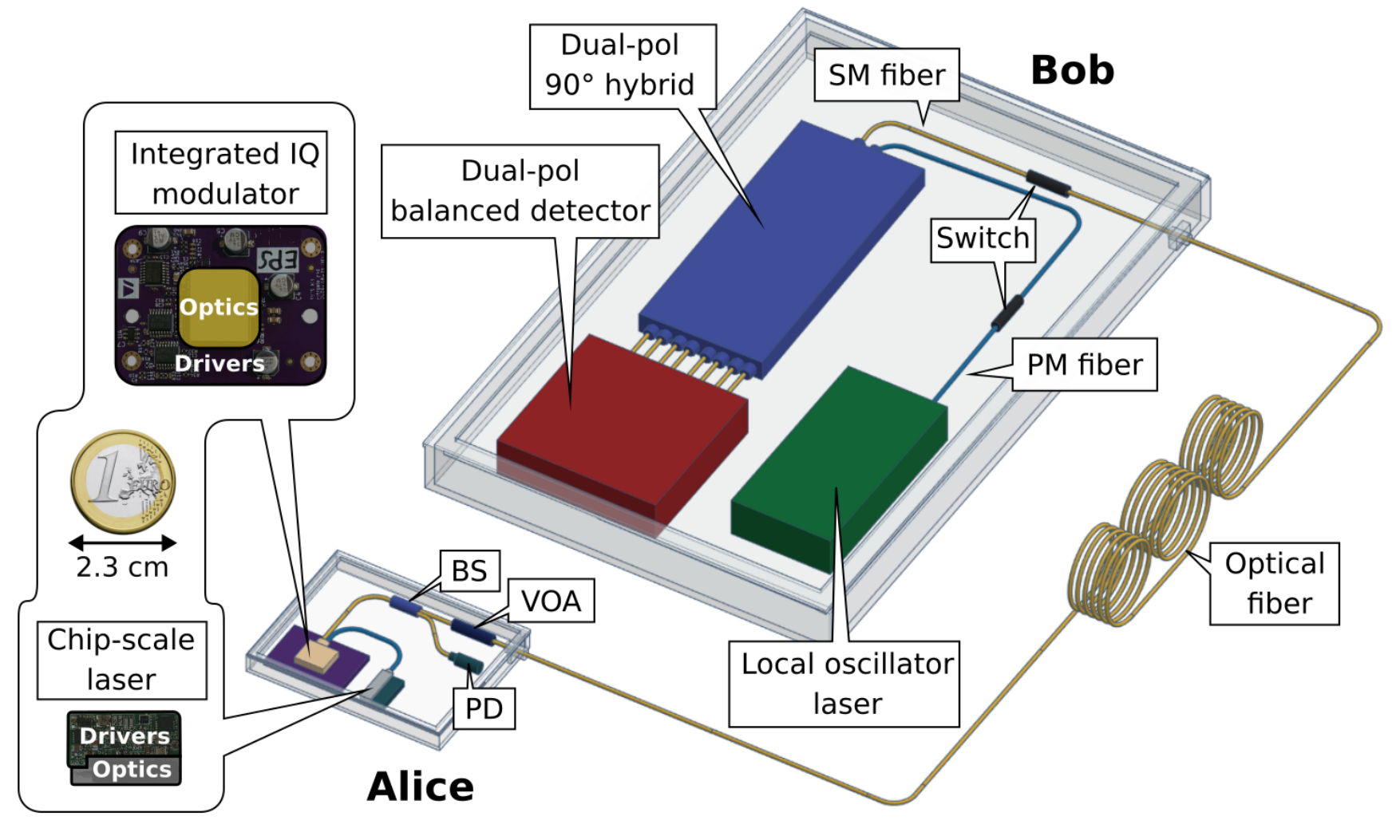}
\caption{Schematic of the experimental setup, to scale. The miniaturized CV-QKD transmitter under study (Alice, bottom left) is connected via optical fiber to the receiver (Bob, top right), which is implemented using bulk optical components. On the left hand side are shown photographs (partially obfuscated) of the chip-scale modules used in this work: the IQ modulator and the laser, shown alongside a 1-euro coin for scale.}
\label{fig:setup}
\label{fig:chip-scale_modules}
\end{figure}

This work builds on a CV-QKD system from the \emph{Andariel Testbed Series}, developed by KEEQuant GmbH.  
The system is composed of two primary blocks: the transmitter (Alice) and the receiver (Bob), which are connected by a single optical fiber. Notably, no additional RF connection e.g. for clock synchronization is needed. The experimental system employed in the study is represented in Fig. \ref{fig:setup}, with the utilized optical components represented to scale.\\
In the present implementation, the main discrete optical components on Alice’s side, the laser and the IQ modulator, have been replaced by chip-scale alternatives. The chip-scale components used in this work are shown in Fig. \ref{fig:chip-scale_modules}, to scale. Compared to the system based on discrete components, the chip-scale IQ modulator and laser achieve volume reductions by factors of 27 and 23, respectively. Considering the entire transmitter, the miniaturized prototype optical components occupy only 5\% of the original volume, reduced from 121 cm$^3$ to 6.7 cm$^3$.

For both nodes, the optical components are interconnected and controlled through dedicated printed circuit boards (PCB). These PCBs host the control electronics, digital-to-analog converters (DACs), analog-to-digital converters (ADCs), and the communication interface with the central processing unit (CPU), providing signal generation, acquisition, and system control.

\subsection{Transmitter}
The transmitter side of the QKD system (Alice) consists of a laser source, an IQ modulator, a VOA, and an optical power meter (PM). The laser generates the coherent optical carrier, which is modulated by the IQ modulator to produce the Gaussian-modulated coherent states required for CV-QKD. The VOA and PM jointly control and monitor the transmitted optical power, thereby setting the desired modulation variance $V_{\text{mod}}$. 

\begin{figure}[]
\centering
\includegraphics[width=\linewidth]{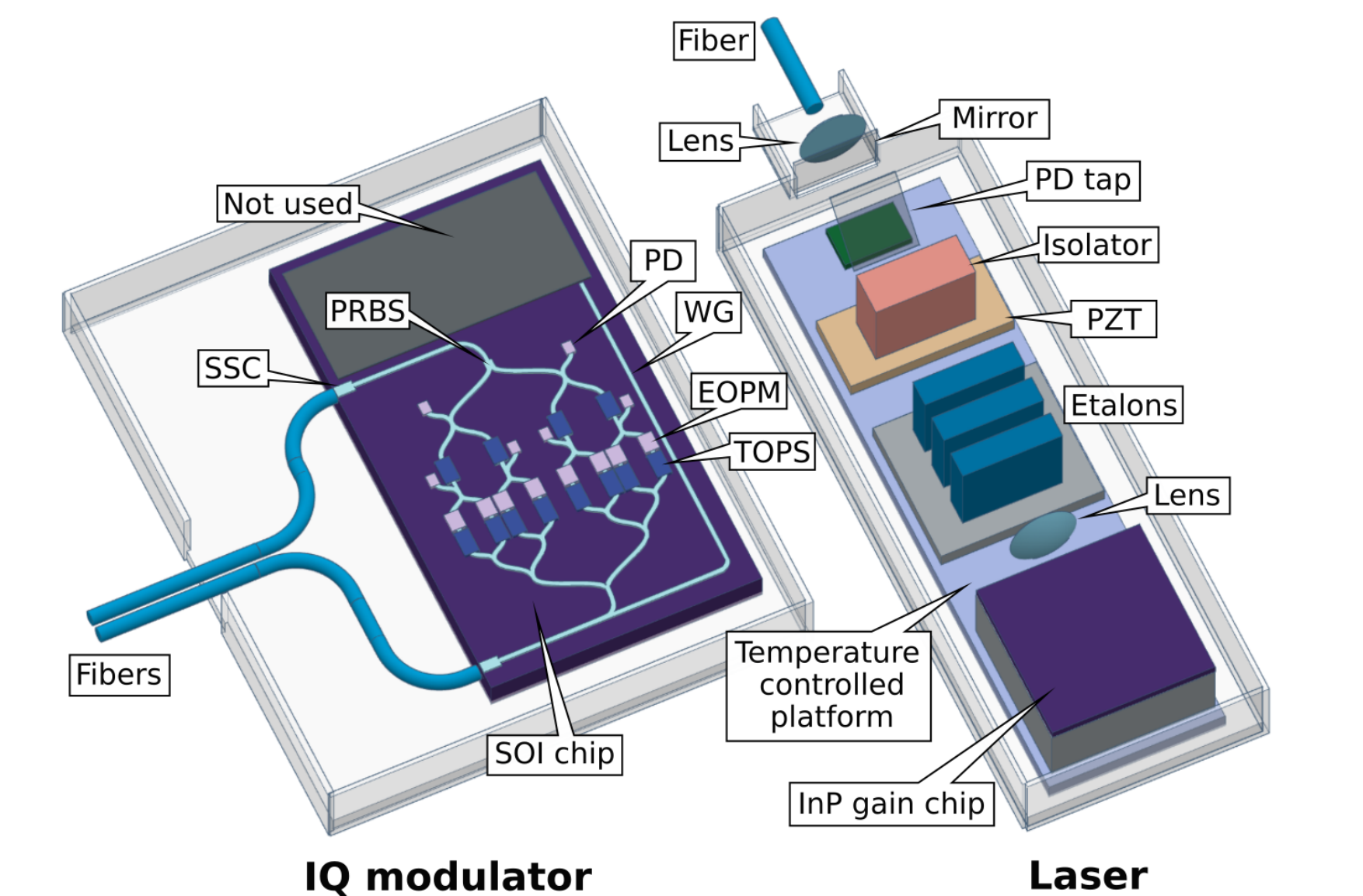}
\caption{Illustration of the schematic of the chip-scale modules. On the left, the PIC incorporating the IQ modulator. Only one polarization was modulated in this study. The details of the unused PIC section are not reported. On the right, the chip-scale laser, composed by the InP gain chip and the external cavity. Components are abbreviated as follows: WG: waveguide; PD: photodiode; TOPS: thermo-optic phase shifter; EOPM: electro-optic phase modulator; PRBS: polarization rotator beam splitter; SSC: spot-size converter. }
\label{fig:chip-scale modules}
\end{figure}

The laser source is a hybrid micro-integrated external-cavity laser (ECL), comprising an InP gain chip coupled to a micro-optical filter cavity. The schematic of the optical module is represented in Fig.~\ref{fig:chip-scale modules}. This configuration preserves the high spectral purity and wide tunability required for CV-QKD, which are often compromised in fully monolithic designs. The driving electronics are co-packaged within the module, ensuring compactness and power-efficiency. The laser achieves a linewidth below 28 kHz, provides full tunability across the C-band, and delivers up to 17 dBm of fiber-coupled optical output power.

The IQ modulation is realized using the modulation section of a telecom-grade packaged PIC, represented in Fig.~\ref{fig:chip-scale modules}. The chip is based on a silicon photonics platform and incorporates a dual-polarization IQ modulator.
The PIC features a single fiber-coupled laser input, half of the input power is directed into the IQ modulator. The optical power is further equally split between the X and Y polarizations.  
The IQ modulator follows a conventional nested Mach–Zehnder structure, consisting of Mach–Zehnder modulators (MZMs) embedded within a larger Mach–Zehnder interferometer (MZI). All the corresponding phase modulators operate in a push–pull configuration. It provides four RF input ports to drive the electro-optic modulators encoding the quadratures $I_x$, $Q_x$, $I_y$, and $Q_y$. In the experiments the signal was modulated on a single polarization. Although the system is designed for high symbol rates (up to 70 GBaud), in this work it is driven at a symbol rate of only 156.25~MBaud.
Thermal optical phase shifters (TOPS) are included to fine-tune the MZM arms for maximum extinction and to maintain the 90° phase bias between the I and Q branches. The measured half-wave voltage is $V_\pi = 0.96$~V. Several on-chip monitoring photodiodes are integrated at the outputs of the MZMs and MZIs to facilitate bias control.  

\subsection{Receiver}
The receiver side of the system (Bob) currently employs discrete optical components. It consists of a narrow-linewidth tunable laser serving as the LO, a dual-polarization 90° optical hybrid, and a balanced dual-polarization detector with low noise transimpedance amplifier (TIA).  
Two optical switches are placed before the signal and LO input ports of the hybrid to enable calibration of the shot-noise unit (SNU).

\section{Real-time CV-QKD Protocol}
CV-QKD allows the two involved parties, Alice and Bob, to establish a shared secret key according to a defined protocol. In our system, we implement the Gaussian-modulated (GM), prepare-and-measure, no-switching protocol \cite{PhysRevLett.93.170504}, which enables real-time operation. The security proof enforced in this work is based on the comprehensive review paper \cite{Laudenbach2018}. The protocol consists of calibration, transmission, and post-processing steps, executed as a series of automated runs. The generated key is stored in a buffer and managed by a Key Management System (KMS), in our case KEEQuant's KMS1, which can forward keys on a network level to provide quantum-secure communication \cite{Horoschenkoff:25}. One protocol shot, i.e. a single run, consists of the following steps:

\begin{enumerate}

\item \textbf{System calibration.} 
Prior to each quantum exchange, the system is calibrated. This includes tuning of hardware components (e.g., laser power stabilization, IQ modulator biasing) and, crucially, calibration of the shot noise unit (SNU), which serves as the reference for the entire protocol. SNU calibration is performed using the standard two-step approach \cite{Laudenbach2018}, where one first measures the electronic noise of the receiver, $N_{\mathrm{el}}$, and then the total noise with no optical input (only LO), $N_{\mathrm{nosig}}$. The SNU is then defined as
\begin{equation}
    N_{\mathrm{shot\ noise}} = N_{\mathrm{nosig}} - N_{\mathrm{el}}.
\end{equation}

\item \textbf{State exchange.} 
Alice prepares Gaussian-modulated coherent states whose quadratures satisfy $p, q \sim \mathcal{N}(0, V_{mod})$. 
The state is sent through the quantum channel and measured by Bob. The end-to-end data flow is described in detail in Section \ref{section:Data processing}.

\item \textbf{Parameter estimation.} 
To assess whether the parties can establish a secure key, the channel parameters (transmittance $\text{T}$ and excess noise $\xi$) are estimated. To do this, Alice reveals a randomly selected subset of her transmitted quantum symbols. Bob uses these to compute the covariance matrix $\Gamma_{AB}$ and from $\text{T}$ and $\xi$.\\
The asymptotic secret key fraction (SKF) is then given by the Devetak-Winter Formula \cite{Laudenbach2018}:
\begin{equation}
\text{SKF} = (1-\nu)(1-\mathrm{FER}) \left( \beta I_{AB} - \chi_{BE} \right),
\label{eq:skf}
\end{equation}
where $I_{AB}$ is the mutual information between Alice and Bob; $\chi_{BE}$ is the Holevo bound in the reverse reconciliation scheme (in the case of individual attacks, it is substituted by the classical information $I_{EB}$) \cite{RevModPhys.81.1301}; $\beta$ is the reconciliation efficiency; $\mathrm{FER}$ is the frame error rate; and $\nu$ is the fraction of disclosed symbols for parameter estimation. If $\text{SKF}>0$, the protocol proceeds; otherwise, it resets to the first step.
We assume a Gaussian attack, since it is proven to be optimal in both individual and collective attacks \cite{PhysRevLett.97.190503}. Under this assumption we retrieve $I_{AB}$ and $\chi_{BE}$ (or $I_{EB}$) via $\text{T}$ and $\xi$ \cite{Laudenbach2018,RevModPhys.81.1301}. We assume the trusted-detector scenario; therefore, the receiver noise $\xi_{\text{rec}}$ and transmittance $\text{T}_{\text{rec}}$ are excluded from the calculation of the Holevo information.
Additionally, the system optimizes the modulation variance $V_{mod}$ based on the estimated channel parameters to maximize the expected key rate for the next quantum exchange.

\item \textbf{Information reconciliation.} 
To convert the measured continuous variables into a bit string, we employ Multi-Dimensional Reconciliation (MDR) \cite{PhysRevA.77.042325}, which maximizes performance at low SNR, in conjunction with Low-Density Parity-Check (LDPC) codes \cite{1057683}, which achieve efficiencies close to the Shannon capacity.

\item \textbf{Confirmation.} 
A 32-bit cyclic redundancy check (CRC32) is performed between Alice and Bob to confirm they hold the same key.

\item \textbf{Privacy amplification.} 
The reconciled keys undergo randomness extraction using a function chosen at random from a family of 2-universal hash function, eliminating any residual information potentially available to an eavesdropper about the final key \cite{Bennett1988PrivacyAB}.

\end{enumerate}

All communication between, Alice and Bob in the steps above is done using cryptographic authentication on the classical public channel. The key material used for that is renewed after each shot and is directly obtained from the protocol's key buffers.

\section{End-to-end Signal Flow}
\label{section:Data processing}
\begin{figure}[]
    \centering
    \includegraphics[width=0.85\linewidth]{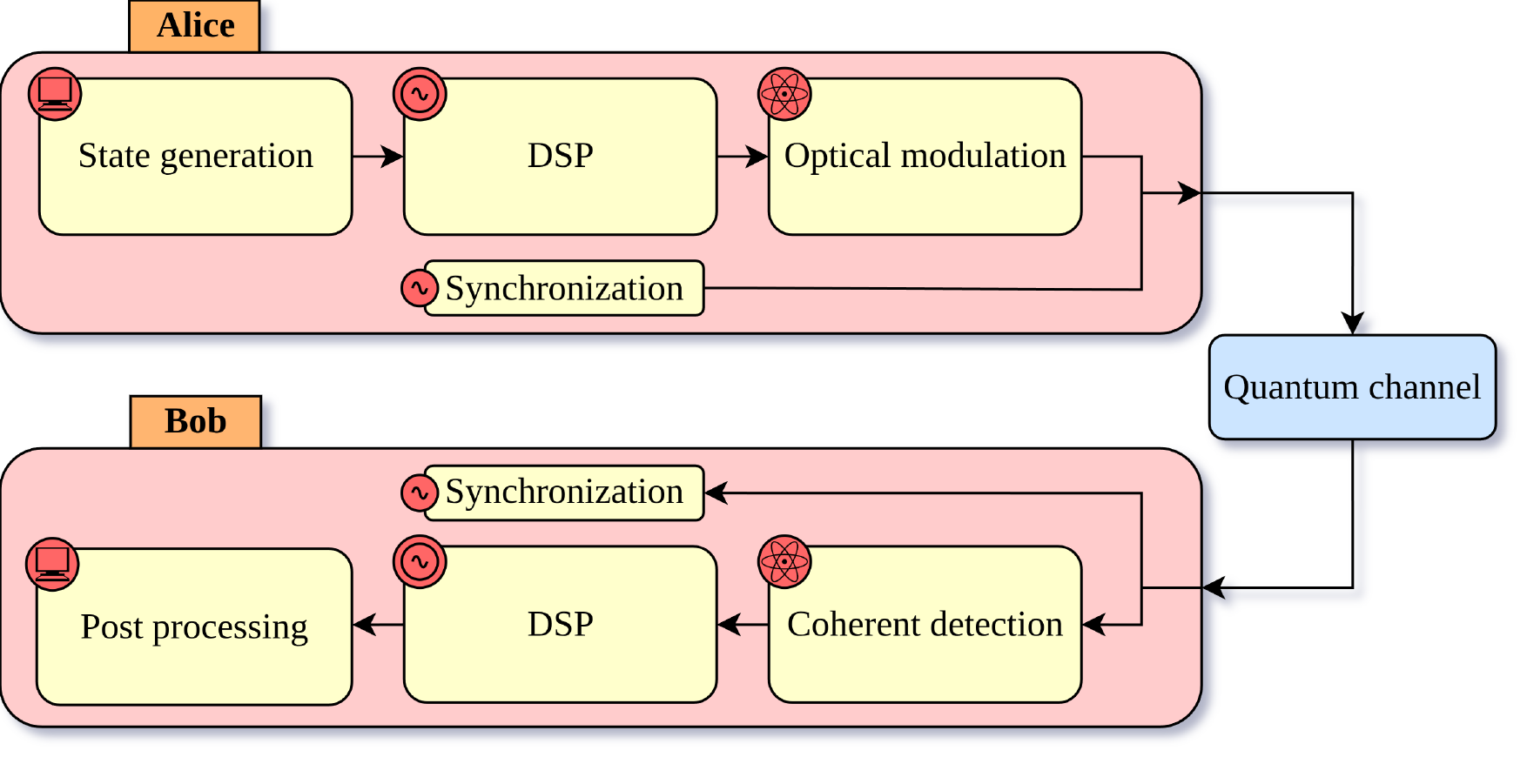}
    \caption{Diagram illustrating the signal flow in the CV-QKD protocol.}
    \label{fig:DSP}
\end{figure}

The exchange of quantum states corresponds to transmission and measurement of Gaussian-modulated quadrature values. To realize this process, digital signals are mapped onto quantum states, transmitted, and then recovered through a series of processing stages. The end-to-end signal flow is illustrated in the flowchart shown in Fig.~\ref{fig:DSP}. 

The relevant steps of the pipeline are the following:
\begin{itemize}

\item \textbf{State generation:}
The signal is generated digitally using a Pseudo-Random Number Generator (PRNG), which produces two independent series of Gaussian-distributed symbols corresponding to the two optical quadratures. 

\item \textbf{Alice DSP:}
The quadrature samples undergo digital upsampling and pulse shaping, confining the spectral content within the available bandwidth. Furthermore, a digital pre-emphasis filter is employed to compensate for the non-flat frequency response of the electrical TX path. To facilitate carrier recovery, the quantum signal is multiplexed with a pilot tone.

\item \textbf{Optical modulation:}
The digital waveform is converted into analog driving voltages for the IQ modulator by means of a DAC. The modulator is biased via Thermo-Optical Phase Shifters (TOPS) to ensure optimal performance. Specifically, the $I$ and $Q$ branches are biased at their minima to suppress the carrier, while the relative phase is stabilized at $90^{\circ}$. Biasing is performed during operation via an automatic control loop. Following modulation, the optical power is monitored via a photodiode. The signal is subsequently attenuated using a VOA to reach the target modulation variance.

\item \textbf{Synchronization:}
The LO is frequency-stabilized with respect to the signal carrier by means of a pilot tone processed within a proportional–integral–derivative (PID) control loop.  
Clock synchronization is achieved through a second synchronization pilot tone, which is used to estimate and compensate for both clock offsets and frequency drifts \cite{Chin2022DigitalSyncCVQKD}.

\item \textbf{Coherent detection:} 
At the receiver, the incoming quantum signal interferes with a LO in a dual-polarization $90^{\circ}$ optical hybrid. The resulting optical outputs are detected by a dual-polarization balanced receiver and digitized by an ADC.

\item \textbf{Bob DSP:}
An equalization filter is implemented to compensate for the frequency response of the receiver. Specifically, a digital Finite Impulse Response (FIR) filter is derived by inverting the experimentally measured transfer function of the receiver. Phase and polarization state recovery are performed using the pilot tone as a reference.

\end{itemize}

The final output of this processing chain consists of the received digital quadrature values, which serve as the data for the subsequent steps of the QKD protocol.

\section{Results}
To evaluate the performance of the CV-QKD system, two separate tests were carried out. In the first test, the system was operated over optical fibers with a total length of 26, 52, 77 and 102 km. In the second test, optical channels with different length were emulated by introducing additional attenuation using a VOA, in order to assess the system performance over increasing channel loss and establish the maximum achievable reach.
Operating in the asymptotic regime under a trusted detector assumption, the protocol is analyzed against both individual and collective attacks. While collective attacks represent the literature benchmark, we also present evaluations under the assumption of individual attacks. This dual approach acknowledges industrial use-cases where maximizing system performance is the driving factor, making a more pragmatic security assumption acceptable given that executing a collective attack exceeds adversarial technical capabilities. The real-time system operation is identical for both scenarios, varying only during privacy amplification, where the key is hashed according to the corresponding secret key fraction. The number of exchanged quantum symbols per shot is $2^{21}$.

\subsection{Real-time CV-QKD Analysis}

In the CV-QKD literature, the reference metric used to evaluate the performance of a CV-QKD system is the secret key rate, which characterizes the key-generation process over time. Previous implementations used an idealized metric that accounted only for the time of the quantum symbols exchange. We will refer to this rate as quantum states exchange key rate ($\text{SKR}_{\text{qse}}$), given by:
\begin{equation}
    \text{SKR}_{\text{qse}} = \frac{\text{n}_{\text{key}}}{\text{t}_{\text{sym}}} = \text{p}_{\text{suc}} \cdot \text{SKF} \cdot \text{f}_{\text{sym}}
\end{equation}
Where $\text{SKF}$ is the secret key fraction, $\text{f}_{\text{sym}}$ is the symbol rate, $\text{n}_{\text{key}}$ is the number of established secret key bits, and $\text{t}_{\text{sym}}$ the time necessary for exchanging the quantum symbols. The shot success rate $\text{p}_{\text{suc}}$ represents the protocol failure factors associated with real-time operation, such as synchronization, parameter estimation, post-processing failures etc.; therefore, it is neglected in offline CV-QKD demonstrations. The reason this metric was commonly adopted lies in typical offline post-processing, which only allows one to infer a rate for the system, since the key is not generated in real time. To do this, the adopted standard is to consider the fundamental time-limiting characteristic of the offline process: the quantum states exchange.
When accounting for the real-time nature of the protocol, this commonly used metric gives only limited information, since it does not reflect the protocol duration.  For this motivation, we suggest to adopt the real-time key rate ($\text{SKR}_{\text{rt}}$), accounting for the entire protocol, including the duration of calibration, post-processing, and other overheads. This metric represents the effective key generated over time that an end-user could access. We define it as follows:
\begin{equation}
\text{SKR}_{\text{rt}} = \frac{\text{n}_{\text{key}}}{\text{t}} = \text{p}_{\text{suc}} \cdot \text{SKF} \cdot \text{f}_{\text{shot}}
\label{eq:skr}
\end{equation}
Here, $\text{n}_{\text{key}}$ is the number of established secret key bits, $\text{t}_{\text{sym}}$ is the total measurement time, and $\text{f}_{\text{shot}}$ represents the shot rate, which in addition to the symbols exchange, incorporates calibration, DSP, synchronization and post-processing time. The two rates are related by:
\begin{equation}
\text{SKR}_{\text{qse}} = \text{SKR}_{\text{rt}} \,(1 + x), \quad 
x = \frac{\text{t} - \text{t}_{\text{sym}}}{\text{t}}.
\end{equation}
They are comparable only when $\text{t} \approx \text{t}_{\text{sym}}$, which represents an ideal scenario, since post-processing remains the primary resource-intensive step.

\subsection{Performance Evaluation over Optical Fiber}

\begin{figure}
    \centering
    \includegraphics[width=\linewidth]{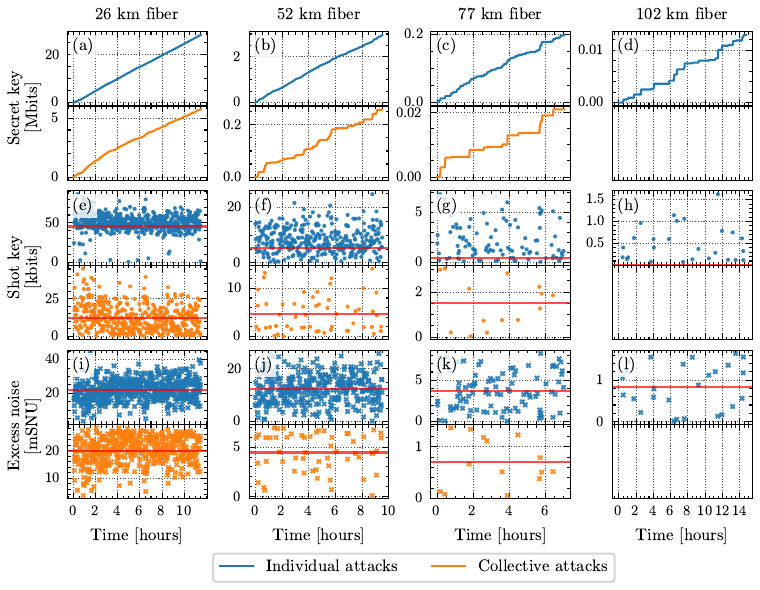}
    \caption{Characterization over time of the CV-QKD system featuring the chip-scale transmitter over optical fibers of 26, 52, 77, and 102 km. The plots show the real-time operation of the protocol, displaying: (a)–(d) the total established secret key; (e)–(h) the key length of the successful shots; and (i)–(l) the corresponding excess noise. The red lines in (e)–(l) for shot key length and excess noise represent their average values. The average shot key length also accounts for the unsuccessful shots.}
    \label{fig:fiber_test}
\end{figure}

\begin{figure}[]
    \begin{subfigure}{0.7\linewidth}
    \includegraphics[width=\linewidth]{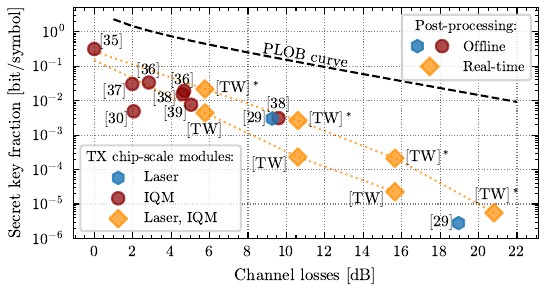}
    \caption{}
    \label{fig:skf_comparison}
    \end{subfigure}
    \begin{subfigure}{0.28\linewidth}
    \includegraphics[width=\linewidth]{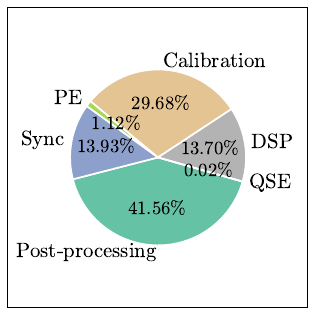}
    \vspace{0.4cm}
    \caption{}
    \label{fig:time_allocation}
    \end{subfigure}
    \caption{(a) Comparison of CV-QKD experimental secret key fraction for transmitters employing chip-scale components versus channel losses. Miniaturized modules and real-time operation are detailed in legend. The dotted orange lines guide the eye for comparison with previous results. [TW] refers this work, $^*$ indicates assumption of individual instead of collective attacks. The dashed black line marks the PLOB bound (Pirandola-Laurenza-Ottaviani-Banchi), fundamental limit for the SKF \cite{Pirandola2017}. 
    (b) Average time allocation of protocol steps for the 52 km fiber measurement, with QSE as the time to exchange the quantum states given the symbol rate, ad PE as parameter estimation duration.}
\end{figure}

The system performance was evaluated over fiber lengths of 26, 52, 77 and 102 km. Measurements where carried out in the lab by means of fiber spools. The corresponding experiments ran continuously for 690, 571, 423, and 881 minutes during which a total of 624, 537, 444 and 1106 shots were acquired, respectively. The characterizing metrics of the CV-QKD system are summarized Table~\ref{tab:fiber}. Fig.~\ref{fig:fiber_test} displays the real-time QKD system performance over the employed optical fibers. The plots showcase the total secret key generated over time, and the underlying single shots, with the corresponding values of generated secret key and excess noise.
\begin{table}[h]
\caption{Evaluation of CV-QKD system parameters over optical fibers of varying lengths. The channel metrics are the time-averaged considering the successful shots. The reported parameters, from left to right, include: fiber length ($\text{L}_{ch}$), real-time key rate ($\text{SKR}_{\text{rt}}$), quantum states exchange key rate ($\text{SKR}_{\text{qse}}$), excess noise and transmittance of channel and receiver ($\xi_{\text{ch}}$,$\text{T}_\text{{ch}}$,$\xi_{\text{rec}}$,$\text{T}_\text{{rec}}$), modulation variance ($\text{V}_{\text{mod}}$), reconciliation efficiency ($\beta$), frame error rate (FER), shot success rate ($\text{p}_{\text{suc}}$), and attack assumption.}
\label{qkd_params_42km_fiber}
\begin{tabu} to \linewidth {X[c] X[c] X[c] X[c] X[c] X[c] X[c] X[c] X[c] X[c] X[c] X[c]}
\toprule
$\text{L}_{\text{ch}}$ & $\text{SKR}_{\text{rt}}$ & $\text{SKR}_{\text{qse}}$ & $\xi_{\text{ch}}$ & $\text{T}_\text{{ch}}$ & $\xi_{\text{rec}}$ & $\text{T}_\text{{rec}}$ & $\text{V}_{\text{mod}}$ & $\beta$ & FER & $\text{p}_{\text{suc}}$ & \multirow{2}{*}{Attack} \\

\small [km] & \small [bit/s]& \small [Mbit/s] & \small [mSNU] & \small [\%] & \small [mSNU] & \small [\%] & \small [SNU] & \small [\%] & \small [\%] & \small [\%] &\\
\midrule
\multirow{2}{*}{26} & \multirow{2}{*}{\shortstack{721 \\ 136}} & \multirow{2}{*}{\shortstack{3.38 \\ 0.69}} & \multirow{2}{*}{\shortstack{22.0 \\ 20.0}}  & \multirow{2}{*}{\shortstack{26.5 \\ 26.5}} & \multirow{2}{*}{\shortstack{141 \\ 141}} & \multirow{2}{*}{\shortstack{48 \\ 48}} & \multirow{2}{*}{\shortstack{4.2 \\ 4.1}} & \multirow{2}{*}{\shortstack{91.2 \\ 91.3}} & \multirow{2}{*}{\shortstack{0.07 \\ 0.02}}  & \multirow{2}{*}{\shortstack{90.5 \\ 73.7}} & \multirow{2}{*}{\shortstack{I.A. \\ C.A.}} \\
 & & & & & & & & & & & \\
\multirow{2}{*}{52} & \multirow{2}{*}{\shortstack{90.5 \\ 7.0}} & \multirow{2}{*}{\shortstack{0.41 \\ 0.036}} & \multirow{2}{*}{\shortstack{12.4 \\ 4.4}} & \multirow{2}{*}{\shortstack{8.7 \\ 8.7}} & \multirow{2}{*}{\shortstack{144 \\ 144}} & \multirow{2}{*}{\shortstack{48 \\ 48}} & \multirow{2}{*}{\shortstack{4.2 \\ 3.4}} & \multirow{2}{*}{\shortstack{92.7 \\ 92.3}} & \multirow{2}{*}{\shortstack{4.1 \\ 0.5}} & \multirow{2}{*}{\shortstack{65.9 \\ 10.1}} & \multirow{2}{*}{\shortstack{I.A. \\ C.A.}} \\
 & & & & & & & & & & & \\
\multirow{2}{*}{77} & \multirow{2}{*}{\shortstack{8.1 \\ 0.6}} & \multirow{2}{*}{\shortstack{0.033 \\ 0.003}} & \multirow{2}{*}{\shortstack{3.7 \\ 0.7}} & \multirow{2}{*}{\shortstack{2.7 \\ 2.7}} & \multirow{2}{*}{\shortstack{144 \\ 144}} & \multirow{2}{*}{\shortstack{48 \\ 48}} & \multirow{2}{*}{\shortstack{7.8 \\ 6.2}} & \multirow{2}{*}{\shortstack{93.7 \\ 92.4}} & \multirow{2}{*}{\shortstack{18.6 \\ 1.8}} & \multirow{2}{*}{\shortstack{21.6 \\ 3.2}} & \multirow{2}{*}{\shortstack{I.A. \\ C.A.}} \\
 & & & & & & & & & & & \\
102 & 0.2 & 0.0009 & 0.8 & 0.8 & 144 & 48 & 13.2 & 94.5  & 31.6 & 2.4 & I.A. \\
\bottomrule
\end{tabu}
\label{tab:fiber}
\end{table}
Over the full measurement periods, the system generated total secret keys under individual / collective attacks of 28.32 Mbit / 5.76 Mbit (at 26 km), 2.96 kbit / 0.26 kbit (at 52 km), 198 kbit / 21 kbit (at 77 km), and 13 kbit / 0 bit (at 102 km). This corresponds to real-time key rates ranging from 721 / 136 bit/s at 26 km, down to 0.2 / 0 bit/s at 102 km, where collective attacks yielded no secret key.
As shown in Fig.~\ref{fig:fiber_test}, the key generation process guaranteed constant key updates over the measurement periods, allowing key-refreshing for real-time encryption. For the shorter measured channel links, key updates were frequent even over short time intervals, ensuring high reliability and low key-update latency, with the few failed shots distributed evenly among synchronization, parameter estimation, and error correction issues. Increasing the channel length we observe a decrease in the success rates, thereby reducing the frequency of key updates. The underlying reason for this reduction was a lower probability of successful parameter estimation, as longer channels impose stricter excess noise constraints. This phenomenon also explains the lower success rate under collective attacks.\\
Regarding noise characterization, the excess noise from successful shots, referred to the channel output, decreased as the fiber length increased from an average value of approximately 20 mSNU to 1 mSNU. A lower average excess noise can be observed in the case of collective attacks, as only an optimal smaller subset of shots meets the requirements to enable secret key generation under a stricter security assumption. The modulated variance was automatically optimized during each shot for secret key fraction maximization, and the average value increased over longer channels from 4 to 13 SNU.\\
Notably, during live operation, the real-time key rate is approximately four orders of magnitude lower than the quantum state exchange key rate. As shown in the time allocation per shot (Fig.~\ref{fig:time_allocation}), the quantum state exchange time, determined by the symbol rate, is negligible compared to the other required steps of the protocol, accounting for just 0.02\% of the shot duration. Instead, the most time-consuming steps are post-processing, calibration, synchronization, and digital signal processing (DSP). Consequently, the quantum state exchange key rate alone provides limited insight for practical deployments, as it fails to capture the primary bottlenecks of CV-QKD systems. For this reason, the real-time key rate serves as a more informative metric; offering a realistic baseline for real-time encryption applications.
Analyzing the protocol time allocation, post-processing currently represents the most severe optimization bottleneck. Its high computational load scales linearly with the number of symbols, counteracting the advantages of higher symbol rates. An efficient implementation on accessible hardware is an ongoing research challenge \cite{9057665, Li2020GPUDecoder, Wang2018HighSpeedErrorCorrection}. 
In contrast, calibration and synchronization overheads do not depend on the number of exchanged symbols. Therefore, their impact can be mitigated by optimizing the number of symbols sent per calibration cycle, factoring the overall weight. Finally, DSP throughput can be enhanced through ASICs or FPGAs implementation, mirroring standard practices in coherent optical communications.

\begin{table}[!ht]
    \centering
    \caption{Comparative table of CV-QKD experimental demonstrations featuring transmitter chip-scale components evaluated over optical fibers. The table lists references, utilized chip-scale components, real-time and quantum states exchange key rate, secret key fraction, channel length and symbol rate.}
    \label{tab:TXcomparison}
    \begin{tabu} to \linewidth {X[c] X[c] X[c] X[c] X[c] X[c]  X[c]}
    \toprule
    \multirow{2}{*}{Ref.} & \multirow{2}{*}{\shortstack{Chip-scale \\ modules}} & $\text{SKR}_{\text{rt}}$ & $\text{SKR}_{\text{qse}}$ & $\text{SKF}$ & $\text{L}_{\text{ch}}$ & $\text{f}_{\text{sym}}$ \\
     & & \small [bit/s] & \small [bit/s] & \small [mbit/sym] & \small [km] & \small [MBaud] \\ \midrule
    \multirow{2}{*}{\cite{Li2023}} & \multirow{2}{*}{\shortstack{Lasers \\ (TX,RX)}} & \multirow{2}{*}{\shortstack{n.a. \\ \scriptsize (offline)}} & \multirow{2}{*}{\shortstack{750k \\ 700}} & \multirow{2}{*}{\shortstack{3.0 \\ 0.003}} & \multirow{2}{*}{\shortstack{50 \\ 100}} & \multirow{2}{*}{250} \\
     & & & & & & \\
    \cite{Aldama2025} & IQM & \multirow{1.5}{*}{\shortstack{n.a. \\ \scriptsize (offline)}} & 78k & 4.9 & 11 & 16 \\
    \cite{Zhang2019} & IQM, ICR & \multirow{1.5}{*}{\shortstack{n.a. \\ \scriptsize (offline)}} & 250k & 313 & 0.002 & 0.8 \\
    \multirow{2}{*}{\cite{Hajomer2025b}} & \multirow{2}{*}{IQM, ICR} & \multirow{2}{*}{\shortstack{n.a. \\ \scriptsize (offline)}} & \multirow{2}{*}{\shortstack{534M \\ 246M}} & \multirow{2}{*}{\shortstack{33.4 \\ 15.4}} & \multirow{2}{*}{\shortstack{10 \\ 20}} & \multirow{2}{*}{16000} \\
     & & & & & & \\
    \cite{Ng2025} & IQM, ICR & \multirow{1.5}{*}{\shortstack{n.a. \\ \scriptsize (offline)}} & 1.21G & 30.3 & 10 & 40000 \\
    \multirow{2}{*}{\cite{Liu2025}} & \multirow{2}{*}{IQM, ICR} & \multirow{2}{*}{\shortstack{n.a. \\ \scriptsize (offline)}} & \multirow{2}{*}{\shortstack{31.1M \\ 5.05M}} & \multirow{2}{*}{\shortstack{19.1 \\ 3.1}} & \multirow{2}{*}{\shortstack{25.8 \\ 50.4}} & \multirow{2}{*}{1625} \\ 
     & & & & & & \\
     \cite{11263038} & TRX & \multirow{1.5}{*}{\shortstack{n.a. \\ \scriptsize (offline)}} & 1.9M & 7.6 & 252.2 & 250 \\ \addlinespace[1ex]
    \multirow{7}{*}{This work} & \multirow{7}{*}{\shortstack{Laser (TX),\\ IQM}} & \multirow{2.4}{*}{\shortstack{721$^*$ \\ 136  \\ \scriptsize (real-time)}} & \multirow{2}{*}{\shortstack{3.38M$^*$ \\ 690k}} & \multirow{2}{*}{\shortstack{21.6$^*$ \\ 4.4}} & \multirow{2}{*}{26} & \multirow{7}{*}{156.25} \\ 
     & & & &  & & \\
    & & \multirow{2.4}{*}{\shortstack{90.5$^*$ \\ 7.0 \\ \scriptsize (real-time)}} & \multirow{2}{*}{\shortstack{410k$^*$ \\ 36k}} & \multirow{2}{*}{\shortstack{2.6$^*$ \\ 0.23}} & \multirow{2}{*}{52} & \\
    & & & &  & & \\
    & & \multirow{2.4}{*}{\shortstack{8.1$^*$ \\ 0.6 \\ \scriptsize (real-time)}} & \multirow{2}{*}{\shortstack{33k$^*$ \\ 3k}} & \multirow{2}{*}{\shortstack{0.21$^*$ \\ 0.02}} & \multirow{2}{*}{77} & \\
    & & & &  & & \\
    & & \multirow{1.5}{*}{\shortstack{0.2$^*$ \\ \scriptsize (real-time)}} & 871$^*$ & 0.006$^*$ & 102 & \\ \addlinespace[1ex]
    \bottomrule
    \end{tabu}
    \footnotesize{Measurements marked by $^*$ assume individual attacks.}
\end{table}

Compared to previous works which attempted to integrate CV-QKD systems on the transmitter side~\cite{Li2023,Aldama2025,Zhang2019,Hajomer2025b,Ng2025,Liu2025,11263038}, the main components of the CV-QKD transmitter—laser and IQ modulator— have been individually integrated on PICs. However, no prior work has demonstrated a transmitter based exclusively on chip-scale components. Moreover, these demonstrations lack real-time operation, a key limitation that we have overcome to bring QKD to operation in real environments. The comparison of secret key fractions across previously reported systems is illustrated in Fig.~\ref{fig:skf_comparison}, with detailed performance metrics summarized in Table~\ref{tab:TXcomparison}. The plot demonstrates that, under collective attacks, our system performs comparably to previous studies in the 5~dB loss region, with a slight decrease in performance in the 10~dB region. This decrease is attributed to the real-time nature of the protocol, which accounts for a reduced success rate. Furthermore, we extend the typical reach reported in earlier studies to a 77~km channel. Under the individual attack assumption, we achieve better performance across the entire range. The distance of 102~km, with attenuation of 20.8~dB, represents the highest reported system reach. These results demonstrate that the chip-scale transmitter delivers excellent performance despite hardware miniaturization. The system proved suitable for real-time operation over typical metropolitan links in a point-to-point architecture. Furthermore, when combined with trusted nodes, it can support long-distance coverage, paving the way for practical and scalable CV-QKD network deployments.

\subsection{Performance Evaluation over Emulated Optical Fiber with Variable Loss}
\begin{figure}
        
    \begin{subfigure}{0.9\linewidth}
    \centering
    \includegraphics[width=\linewidth]{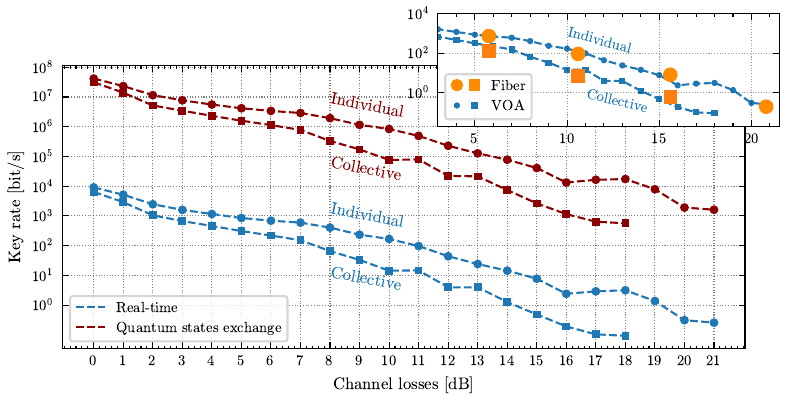}
    \caption{}
    \label{fig:keyrate}
    \end{subfigure}

    \begin{subfigure}{0.35\linewidth}
    \centering
    \includegraphics[width=\linewidth]{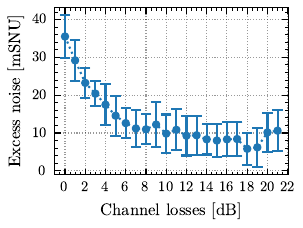}
    \caption{}
    \label{fig:noise}
    \end{subfigure}
    \centering
    \begin{subfigure}{0.35\linewidth}
    \centering
    \includegraphics[width=\linewidth]{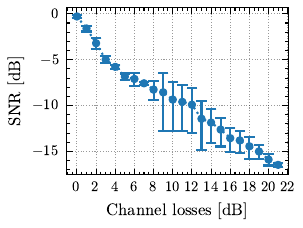}
    \caption{}
    \label{fig:snr}
    \end{subfigure}
    
    \caption{CV-QKD system characterization over quantum channels with variable losses. The plots represent relevant metrics vs. channel losses, specifically: (a) shows real-time and quantum states exchange key rate, as well as a close-up for comparison of real-time key rate over emulated and real fiber. (b) and (c) show average and standard deviation of channel excess noise and quantum symbols SNR at the receiver. Values include unsuccessful shots, excluding outliers. (b) shows the average time allocation of protocol steps for the 52 km fiber measurement, with QSE as the time to exchange the quantum states given the symbol rate, ad PE as parameter estimation duration.}
    \label{fig:reach_test}
\end{figure}

The second measurement concerns the evaluation of the system performance over a quantum channel with controlled channel losses, implemented via signal attenuation using a VOA. The experiment consisted of a total of 1872 shots acquired over 29 hours, while sweeping the channel losses from 0 dB (back-to-back configuration) up to 21 dB. These values correspond to emulated fiber lengths ranging from 0 to 105 km, assuming typical fiber attenuation at 1550~nm of 0.2~dB/km. The results are reported in Fig.~\ref{fig:reach_test}.

The system enabled key generation up to channel losses of 18 dB under collective attacks and 21 dB under individual attacks. At back-to-back configuration, the real-time key rates were 6.4 kbit/s and 9.1 kbit/s, dropping to 0.1 bit/s and 0.3 bit/s at maximum reach. These results correspond to quantum state exchange rates ranging from 31.3 Mbit/s to 1.6 kbit/s for collective attacks, and 41.6 Mbit/s to 0.6 kbit/s for individual attacks.These results are shown in Fig.~\ref{fig:keyrate}. As we emulated fiber propagation via VOA attenuation, the dominant noise contribution must originate from the transmitter, which includes the laser (phase noise, RIN) and the IQ modulator (extinction ratio, compression, drivers). Observing the excess noise, referred to the channel output, including successful and unsuccessful shots, as shown in Fig.~\ref{fig:noise}, we would expect the intrinsic noise, given by the back-to-back measurement (36 mSNU), to scale with the channel losses. This phenomenon is initially observed, but, interestingly, the excess noise reaches a plateau around 10 mSNU when approaching high channel losses, degrading the performance. Phase recovery, achieved via a pilot tone, is the main contributor to this behavior, in fact the pilot is subjected to the same attenuation of the symbols, both when $\text{V}_{\text{mod}}$ is set and during propagation through the channel. As shown in Fig.~\ref{fig:snr}, the symbols signal-to-noise ratio (SNR) degrades by 17 dB over evaluated losses range, the same happens to the pilot tone. At high channel losses, the pilot-tone SNR degrades to the point of affecting phase-recovery accuracy, resulting in additional artificial excess noise, which explains the reduced shot success rate, given the tighter key-generation requirements for the channels parameters.\\
When we compare the experimental results obtained over real fiber with those from the emulated optical channel, as shown in the top right region of Fig.~\ref{fig:reach_test}a, we observe consistency in the key rate datapoints, suggesting that the behaviour of the optical link can be effectively emulated by VOA attenuation. Propagation effects had no observable effect on the channel parameters, as at such low power (in the pW regime), and bandwidth (in the order of $10^8$ Hz), non-linear effects and dispersion penalties are negligible. Based on these measurements, we establish a maximum system reach of 21 dB for individual attacks and 18 dB for collective attacks. For individual attacks, the 21 dB limit confirms the prior measurements over the 102 km optical fiber. Conversely, for collective attacks, the 18 dB threshold demonstrates an extended system reach corresponding to 90 km of standard fiber, improving over our 77 km fiber measurement.

\section{Conclusions}
In this work, we demonstrated the first CV-QKD system with a transmitter composed exclusively of chip-scale components. The presented TX module leverages telecom-grade chip-scale IQ modulator and laser modules, adapted for CV-QKD applications. This miniaturized design enables a significant reduction in footprint, cost and manufacturing scalability, allowing a complete CV-QKD transmitter to be implemented in currently precluded form factors, such as pluggables or PCIe cards. Equipped with this module, the system generates real-time secure keys under individual and collective attacks up to 9.1 kbit/s and 6.4 kbit/s back-to-back, and 0.3 and 0.1 bit/s at their maximum reach, corresponding to losses of 21 dB and 18 dB, equivalent to 105 and 90 km fibers. This performance enables secure communication over typical metropolitan links and can extend to longer distances when combined with trusted nodes, providing key rates compatible with real-time AES encryption. These results represent a significant step towards practical, compact, and deployable quantum-secured communication systems, paving the way for broader applications in finance, healthcare, and beyond.\\
Future research will focus on extending this level of miniaturization to the receiver, incorporating chip-scale laser and integrated coherent detector, to demonstrate a complete scalable CV-QKD system. Additional efforts will target DSP and post-processing optimization to further increase system reach and real-time key rates.

\small

\paragraph{Funding:}
Funded by the European Union as part of the projects QuNEST and SEQRET (Grant Agreements: 101120422, 101091591). Views and opinions expressed are, however, those of the author(s) only and do not necessarily reflect those of the European Union or REA. Neither the European Union nor the granting authority can be held responsible for them.

\paragraph{Acknowledgments:} The authors thank Laurent Schmalen and Frank Volland for their valuable comments and for reviewing the manuscript, and the whole KEEQuant team for insightful discussions and support.

\paragraph{Disclosures:}
I.S.: KEEQuant GmbH (E); M.H.: KEEQuant GmbH (E); M.B.: KEEQuant GmbH (E); E.S.: KEEQuant GmbH (E,I); P.G.: KEEQuant GmbH (E); U.E.: KEEQuant GmbH (E,I); E.E.: KEEQuant GmbH (E,I); I.K.: KEEQuant GmbH (E,I). S.R. declares no conflicts of interest.

\paragraph{Data availability:}
Data underlying the results presented in this paper are not publicly available at this time but may be obtained from the authors upon reasonable request.

\bibliography{bibliography}

\end{document}